\documentclass[prd,aps,showpacs,nofootinbib,preprint,tightenlines]{revtex4}
\usepackage{amssymb}
\usepackage{graphicx,color}
\usepackage{bm}

\newcommand{\checked}[1]{}

\usepackage{mathrsfs}

\usepackage{amsmath}

\usepackage{amsthm}

\newcommand{\beq}{\begin{equation}}
\newcommand{\eeq}{\end{equation}}
\newcommand{\bqa}{\begin{eqnarray}}
\newcommand{\eqa}{\end{eqnarray}}

\begin{document}

\title{Flow Effects on Jet Energy Loss with Detailed Balance}


\author{Luan Cheng$^{a,b,c}$, Jia Liu$^{b,c,d}$ and Enke Wang$^{b,c}$}
\affiliation{$^a$Institute of Theoretical Physics, Dalian University
of Technology, Dalian 116024, China\\ $^b$Institute of Particle
Physics, Central China Normal
University, Wuhan 430079, China\\
$^c$Key Laboratory of Quark $\&$ Lepton Physics (Central China Normal\\
University), Ministry of Education, China \\
$^d$Department of Physics, The Ohio State University, Columbus, OH 43214, USA}

\begin{abstract}
In the presence of collective flow a new model potential describing
the interaction of the hard jet with scattering centers is derived
based on the static color-screened Yukawa potential. The flow effect
on jet quenching with detailed balance is investigated in pQCD. It
turns out, considering the collective flow with velocity $v_z$ along the jet
direction, the collective flow decreases the LPM destructive
interference comparing to that in the static medium. The gluon
absorption plays a more important role in the moving medium.
 The collective flow increases the energy gain from gluon absorption, however,
decreases the energy loss from gluon radiation, which is $(1 - v_z )$ times as
that in the static medium to the first order of opacity. In the presence of collective flow,
the second order in opacity correction is relatively small compared to the first order. So that the
total effective energy loss is decreased. The flow dependence of the
energy loss will affect the suppression of high $p_T$ hadron
spectrum and anisotropy parameter $v_2$ in high-energy heavy-ion
collisions.
\end{abstract}

\maketitle


\section{Introduction}

One of the most striking features of nucleus-nucleus collisions at
the Relativistic Heavy Ion Collider (RHIC) is the collective flow.
In recent years this phenomenon has been a subject of intensive
theoretical and experimental
studies\cite{Lorstad,Voloshin1,Wiedemann1,Voloshin2}. It is believed
that the medium produced in nucleus-nucleus collsions at RHIC
equilibrates efficiently and builds up a flow field.

Gluon radiation induced by multiple scattering of an energetic
parton propagating in a dense medium leads to induced parton energy
loss or jet quenching. As discovered in high-energy heavy-ion
collisions, jet quenching is manifested in both the
suppression of single inclusive hadron spectrum at high transverse
momentum $p_T$ region\cite{phenix} and the disappearance of the
typical back-to-back jet structure in dihadron
correlations\cite{star}. Extentive theoretical investigation of jet
quenching has been widely carried out in recent
years\cite{GW94,BDPMS,Zakharov,GLV,Wiedua,GuoW}. Most of the jet
quenching theory research are studied in a static medium based on
the static color-screened Yukawa  potential proposed by Gyulassy and
Wang\cite{GW94}.
However, the medium is not static, the collective flow need to be
considered\cite{BDMS,SW,WW}.
Later, the interaction between the jet and the target partons in the
presence of collective flow was modeled by a momentum shift
$\mathbf{q}_0$ perpendicular to the jet direction in the
Gyulassy-Wang's static potential\cite{ASWiedemann}, but this
assumption lacks sufficient theoretical evidence. Afterwards, local
transport coefficient $\hat{q}$, which is related to the squared
average transverse momentum transfer from the medium to the hard
parton per unit length, has been investigated in the presence of
transverse flow\cite{BMS,Liu}. By applying a Lorentz boost of $\hat{q}$,
a macroscopic result for parton energy loss based on BDMPS calculation\cite{BDPMS}
was investigated with assuming the hard parton travels with light velocity.
However, this method is adaptable only for light quark energy loss.
Heavy quark energy loss can not be studied in this way since its velocity is less than c.
So a method to study both light and heavy quark energy loss in
 the presence of collective flow is needed. PQCD, a theory describing the strong interactions
 between quarks and gluons, can be a good selection here.
  Moreover, the most interesting feature of the result of jet quenching is the
quadratical distance dependence of the total energy loss because of
the non-Abelian nature of QCD radiation and the
Landau-Pomercanchuk-Migdal (LPM) interference. Both the non-Abelian
feature and LPM effect were studied in pQCD\cite{GW94,BDPMS,GLV}. So
that it is an interesting issue to study jet quenching in the
presence of collective flow in pQCD.
In addition, only
radiative energy loss can be studied in Ref.\cite{BMS,Liu}, the
detailed balance effect with gluon absorption, which need to be
studied in pQCD, cannot be included. It has been shown that the
gluon absorption plays an important role for the intermediate jet
energy region\cite{Wang:2001sf}.

 In this letter, we report a
first study of the parton energy loss with detailed balance in the
presence of collective flow in pQCD. We first determine the model
potential to describe the interaction between the energetic jet and
the scattering target partons of the moving
quark-gluon medium using Lorentz boosts. Based on this new
potential, we then consider both the radiation and absorption
induced by the self-quenching and multiple scattering to the first order
 in opacity in the presence of flow. We obtains to the zeroth order
opacity, the energy loss is dominated by the final-state thermal
absorption. This thermal absorption is $(1+v_z)^2$ times as that in
the static medium because of different gluon distribution in the
moving medium. To the first order in opacity
the collective flow changes the LPM destructive interference. It
reduces the jet energy loss induced by rescattering with stimulated
emission depending on the direction of the flow in the positive jet
direction, but hardly changes thermal absorption by rescattering.
In the framework of opacity expansion developed by
Gyulassy, L\'{e}vai and Vitev (GLV)\cite{GLV} and
Wiedemann\cite{Wiedua}, it was shown that the higher order
corrections contribute little to the radiative energy loss in the static medium.
Here in the moving medium with collective flow. we will calculate the contribution
from the second order in opacity to the energy loss.  We are led to the conclusion that in the presence of flow,
the second order in opacity correction also contributes little as in the static case.
Overall, the collective flow increases energy gain, but decreases
emitted energy loss, so that the total effective energy loss is
decreased.

\section{The Potential Model}

To calculate the induced radiation energy loss of jet in a static
medium, the interaction potential is assumed in the Gyulassy-Wang's
static model \cite{GW94} that the quark-gluon medium can be modeled
by N well-separated color screened Yukawa potentials,
\begin{eqnarray}
V_i^a(q_i)=2\pi\delta(q_i^0)\frac{4\pi\alpha_s}
{\mathbf{q}^2+\mu^2}e^{-i\mathbf{q}\cdot \mathbf{x_i}}
T_{a_i}(j)T_{a_i}(i) \label{GWpotential}\, ,
\end{eqnarray}
where $\mu$ is the Debye screening mass, $\alpha_{s}=g^2/4\pi$ is strong coupling
constant, $T_{a_i}(j)$ and $T_{a_i}(i)$ are the color matrices for the jet and
target parton at $\mathbf{x}_i$ respectively. In this potential, each scattering has no energy
transfer ($q_i^0=0$) but only a small momentum $\mathbf{q}$ transfer
with the medium. If using the four-vector potential, the
Gyulassy-Wang's static potential can be denoted as
$A^{\mu}=(V_i^a(q_i),\mathbf{A}(q_i)=0)$.

As is well known in Electrodynamics, the static charge produces a
static Coulomb electric field, while a moving charge produces both
electric and magnetic field.  In analogy a moving target parton in
the quark-gluon medium will produce color-electric and
color-magnetic fields simutaneously due to the collective flow.
Therefore, the static potential model should be reconsidered.

\begin{figure}
\begin{center}
\includegraphics[width=63mm]{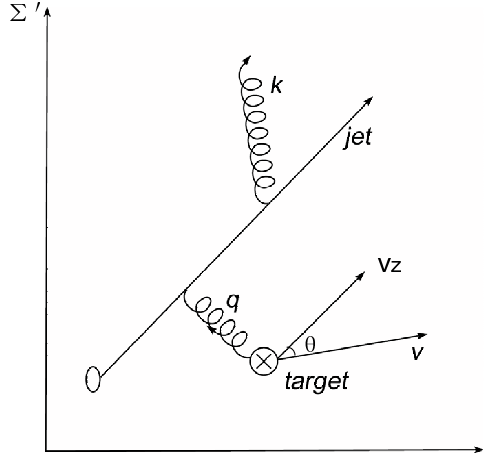}
\end{center}
\vspace{-12pt} \caption{\label{fig:pic1} View of kinematics in the
quark-gluon medium with collective flow. }
\end{figure}

The purpose of this paper is to study the different energy loss of
the same jet in a static medium and the quark-gluon medium with
collective flow. The moving medium with a velocity $\mathbf{v}$ can
be obtained through Lorentz boost of the rest frame as illustrated
in Fig.\ref{fig:pic1}. However, since we consider energy loss of the same jet,
the jet momentum need not been Lorentz transformed. So we first take
a Lorentz transformation for exchanged four-momentum $q$, and then
for four-vector potential $A^{\mu}$, we can then write
$A^{\mu}=(V_{i(flow)}^a(q_i),\mathbf{A}_{(flow)}(q_i))$ in the
observer's system frame $\Sigma'$ as
\begin{equation}
\left\{
\begin{array}{ll}
V_{i(flow)}^a(q_i){=}2\pi\delta(q_i^0{-}\mathbf{v}\cdot\mathbf{q})
e^{-i\mathbf{q}\cdot
\mathbf{x}_i} \tilde{v}(\mathbf{q})T_{a_i}(j)T_{a_i}(i)\, , \\
\mathbf{A}_{(flow)}(q_i){=}2\pi\delta(q_i^0{-}\mathbf{v}\cdot\mathbf{q})
\mathbf{v} e^{-i\mathbf{q}\cdot
\mathbf{x}_i}\tilde{v}(\mathbf{q})T_{a_i}(j)T_{a_i}(i)\, ,
\end{array}
\right.
 \label{flowpotential}
\end{equation}
where
$\tilde{v}(\mathbf{q})={4\pi\alpha_s}/({\mathbf{q}^2-(\mathbf{v}\cdot\mathbf{q})^2+\mu^2})$.
The new potential differs from Gyulassy-Wang's static potential in
that the collective flow of the quark-gluon medium produces a
color-magnetic field and the flow leading to non-zero energy
transfer $q_i^0=\mathbf{v}\cdot\mathbf{q}$, which will affect jet
energy loss as we will show below.

Elastic cross section for small transverse momentum transfer between
jet and target partons can be deduced as
\begin{equation}
\frac{d\sigma_{el}}{d^2 \mathbf{q}}=\frac{C_RC_2}{d_A}\frac{\mid
\tilde{v}(\mathbf{q}) \mid ^2 }{(2\pi)^2}\, , \label{elasticcrosec}
\end{equation}
where $C_R$ and $C_2$ are the Casimir of jet and target parton in
fundamental representation in $d_R$ dimension, respectively. $d_A$
is the dimension of corresponding adjoint representation. Our result
agrees with the GLV elastic cross section in static potential when
the flow velocity goes to zero \cite{GW94}.

\section{Flow Effect on Effective Energy Loss With Absorption}

\subsection{Energy Loss to Zeroth Order in Opacity}
At zeroth order in opacity, the jet has no interaction with the
target parton, we obtain the same factorized radiation amplitude off
a quark
 \begin{equation}
 \label{rad0}
   R^{(0)}=2ig T_c
   \frac{{\bf k}_{\perp}\cdot{\mathbf \epsilon}_{\perp}}
   {{\bf k}^2_{\perp}}\, ,
 \end{equation}
as that in the static medium in Ref. \cite{Wang:2001sf}, where $k=[2\omega,{k}_{\perp}^2/{2\omega},
  \mathbf{k}_{\perp}]$ is the four momentum of the radiated gluon with polarization $\epsilon(k)=[0,\mathbf{\epsilon}_{\perp}\cdot
 \mathbf{k}_{\perp}/{\omega},\mathbf{\epsilon}_{\perp}]$, $T_c$
is the color matrix, and $\alpha_s=g^2/4\pi$ is the strong coupling constant.

Taking account of stimulated emission and thermal absorption in a thermal medium
with temperature $T$, subtracting the gluon radiation spectrum in the vacuum, one obtains
the energy loss to the zeroth order in opacity,
\begin{eqnarray}
 {\Delta E^{(0)}_{abs}}&=&\frac{\alpha_s C_R}{2\pi}E\int dx \int \frac{d{\bf k}_{\perp}^2}{{\bf k}_{\perp}^2}[-P(-x)N_g(xE)+P(x)N_g(xE)\theta(1-x)]\,
 \nonumber \\
 &{\approx}& {-}(1+v_z)^2
   \frac{\pi\alpha_s C_R}{3}
   \frac{T^2}{E}\left[
     \ln\frac{4ET}{\mu^2}{+}2{-}\gamma_{\rm E}
     {+}\frac{6\zeta^\prime(2)}{\pi^2}\right],
 \label{elossab00}
\end{eqnarray}
where $x=\omega/E$, $E$ is the initial jet energy. Here $N_g(|{\bf k}|)=1/[\exp({k}\cdot
u/T)-1]\approx 1/[\exp((1-v_z)\cdot|{\bf k}|/T)-1]$ is the thermal
gluon distribution, $u$ is the four velocity of the collective
flow, $v_z$ is the flow velocity along jet direction, the splitting
function $P_{gq}(x)\equiv{P(x)}/{x}=[1+(1-x)^2]/{x}$ for
$q\rightarrow qg$. In Eq.(\ref{elossab00}), $\gamma_{\rm E}\approx
0.5772$, $\zeta^\prime(2)\approx -0.9376$ and $T$ is the thermal
finite temperature. Although the radiation amplitude is the same as
that in the static case, the energy gain changes to $(1+v_z)^2$
times as that in the static medium because the collective flow
changes the thermal gluon distribution function.

\subsection{Energy Loss to the First Order in Opacity}

When a hard parton goes through the quark-gluon medium, it will
suffer multiple scattering with the parton target inside the medium. Consider the hard parton produced at
$\tilde{x}_0=(z_0,\mathbf{x}_{\perp})$ with initial energy $E$. The hard parton
interacts with the target parton at
$\tilde{x}_1=(z_1,\mathbf{x}_{1\perp})$ with flow velocity
$\mathbf{v}$ by exchanging gluon with four-momentum $q$,  and
emerges with final four-momentum $p$. In the light-cone components,
\begin{eqnarray}
p = [E^++2\mathbf{v}\cdot
\mathbf{q},\mathbf{p}_-,\mathbf{p}_{\perp}^+]\, ,
 \label{lightconecom}
\end{eqnarray}
where $E^+=2E\gg\mu$.

At first order in opacity, consider the jet has the simplest case of
elastic interactions with an array of the new potentials Eq.(\ref{flowpotential}) localized at $\tilde{x_i}=(z_i,\mathbf{b}_i)$ using,
 \begin{equation}
 \label{HIt}
   H_I(t)=\int d^3\overrightarrow{\mathbf{x}}\sum_{i=1}^N A(\overrightarrow{\mathbf{x}}-\overrightarrow{\mathbf{x}}_i) T_a(i)\phi^{\dag}(\overrightarrow{\mathbf{x}},t)T_a(R)\hat{D}(t)\phi(\overrightarrow{\mathbf{x}},t)\, ,
 \end{equation}
 where $A(\overrightarrow{\mathbf{x}}-\overrightarrow{\mathbf{x}}_i)$ is the potential we modeled with considering flow, $\hat{D}(t)=i\overleftrightarrow{\partial_t}$ and $Tr T_a(i)T_b(j)=\delta_{ij}\delta_{ab}C_2(T)d_T/d_A$.

So the scattering amplitude with one of the target partons, as compared to the static case, is changed to
 \begin{eqnarray}
 \label{rad1}
   M^{(1)} &\propto& -i(2p-q)_{\mu}A^{\mu}(q) \nonumber \\
           &\propto& 2\pi\delta(q_i^0{-}\mathbf{v}\cdot\mathbf{q})
e^{-i\mathbf{q}\cdot \mathbf{x}_i}
T_{a_i}(j)T_{a_i}(i)(2E+\mathbf{v}\cdot \mathbf{q})R^{(1)}
           \, .
 \end{eqnarray}
With considering flow, the four-vector potential $A^{\mu}(q)$ replaces the static potential $v(q)$ \cite{GW94} in the static case.  One then obtains that the factorized amplitude $R^{(1)}=(1-v_z)\tilde{v}(\mathbf{q})$ is changed by
 the collective flow with a factor $(1-v_z)$.

Here we investigate the rescattering-induced radiation to the first order in opacity with
considering the flow effect resulting from the moving parton target. Based on our new potential in Eq.(\ref{flowpotential}) by
considering collective flow of the quark-gluon medium, we obtain the factorized
radiation amplitude associated with a single rescattering,
\begin{eqnarray}
R^{(S_1)}&=&2ig\Bigl(\mathbf{H}T_aT_c+\mathbf{B_1}e^{\frac{i\omega_{0}z_{10}}
{1-v_z}}[T_c,T_a] -2v_zT_aT_c
\mathbf{H}(1-e^{\frac{i\omega_{0}z_{10}}{1-v_z}})
+\mathbf{C_1}e^{\frac{i(\omega_{0}-\omega_{1})z_{10}}{1-v_z}}
  [T_c,T_a] \Bigr)\cdot {\mathbf \epsilon}_{\perp}\nonumber \\
  && \times(1-v_z)\, ,
\label{singleamp}
\end{eqnarray}
where $z_{10}=z_1-z_0$,
\begin{eqnarray}
 \label{omega}
   \omega_0&=&\frac{{\bf k}_{\perp}^2}{2\omega}\, ,
   \quad
   \omega_1=\frac{({\bf k}_{\perp}{-}{\bf q}_{\perp})^2}
      {2\omega}\, ,
 \\
   {\bf H}&=&\frac{{\bf k}_{\perp}}{{\bf k}_{\perp}^2}\, ,
   \quad
   {\bf C}_1=\frac{{\bf k}_{\perp}{-}{\bf q}_{\perp}}
      {({{\bf k}_{\perp}{-}{\bf q}_{\perp}})^2}\, ,
   \quad
   {\bf B}_1={\bf H}{-}{\bf C}_1\, .
 \end{eqnarray}
Different from the static medium case, the single rescattering
amplitude depends on the collective flow of the quark-gluon medium.

The interference between the process of double scattering and no
rescattering should also be taken into account to the first order in
opacity \cite{GLV}. Assuming no color correlation between different
targets, the double rescattering corresponds to the ``contact
limit'' of double Born scattering with the same target \cite{GLV}.
Assuming the flow velocity $|\mathbf{v}|\ll 1$, with our new
potential in Eq.(\ref{flowpotential}) the radiation amplitude can be
expressed as
\begin{eqnarray}
 \label{ampd}
   R^{(D_1)}&=&2ig T_c e^{\frac{i\omega_0z_{10}}{1-v_z}}
   \Bigl(-{{C_R+C_A}\over 2}{\bf H} e^{-\frac{i\omega_0z_{10}}
   {1-v_z}}
    +{C_A\over 2}{\bf B}_1
    +2v_z\frac{C_R-C_A}{2}{\bf H}(1-e^{-\frac{i\omega_0z_{10}}{1-v_z}})
 \nonumber \\
    &&
     +{C_A\over 2}{\bf C}_1 e^{-\frac{i\omega_1z_{10}}{1-v_z}}\Bigr)
     \cdot {\mathbf \epsilon}_{\perp}(1-v_z)^2,
\end{eqnarray}
where $C_A$ is the Casimir of the target parton in adjoint
representation in $d_A$ dimension. The double rescattering amplitude
also depends on the collective flow.

To the first order in opacity, we then derive the induced radiation
probability including both the stimulated emission and thermal
absorption as
\begin{eqnarray}
 \label{probafirstord}
   {{dP^{(1)}}\over d\omega}&=&{{C_2}\over {8\pi d_A d_R}}{N\over A_{\perp}}
     \int {{dx}\over x}  \int
      {{d^2{\bf k}_{\perp}}\over {(2\pi)^2}}
     \int {{d^2{\bf q}_{\perp}}\over {(2\pi)^2}} P({\omega\over E}) |R^{(1)}|^2\left\langle
      Tr\left[|R^{(S_1)}|^2{+}2
      Re\left(R^{(0)\dagger}R^{(D_1)}\right)\right]\right\rangle
     \nonumber\\
     &&
     \Big[\left(1{+}N_g(xE)\right)\delta(\omega{-}xE)\theta(1{-}x)
      {+}N_g(xE)\delta(\omega{+}xE)\Big]
      \nonumber \\
   &\approx & {{\alpha_s C_2 C_R C_A}\over {d_A \pi}}{N\over A_{\perp}}
     \int {{dx}\over x}  \int
      {{d{\bf k}_{\perp}^2}\over {\bf k}_{\perp}^2}
     \int {{d^2{\bf q}_{\perp}}\over {(2\pi)^2}} P({\omega\over
     E})(1-v_z)^2v^2({\bf q}_{\perp}){{{\bf k}_{\perp}\cdot {\bf q}_{\perp}} \over
      {\left({\bf k}_{\perp} - {\bf q}_{\perp}\right)^2}}
       \left\langle Re(1{-}e^{\frac{i\omega_1z_{10}}{1{-}v_z}})\right\rangle
      \nonumber \\
        &&
         \Big[\left(1{+}N_g(xE)\right)\delta(\omega{-}xE)\theta(1{-}x)
         +N_g(xE)\delta(\omega+xE)\Big]\, ,
\end{eqnarray}
where $v({\bf q}_{\perp})={{4\pi\alpha_s}/ ({{\bf
q}^2_{\perp}+\mu^2}})$.

The non-Abelian LPM effect is seen in Eq.(\ref{probafirstord}) as
arising from the gluon formation factor
$1-\exp(i\omega_1z_{10}/(1-v_z))$. The formation time of gluon
radiation $\tau_f\equiv (1-v_z)/\omega_1$ becomes shorter (longer),
the LPM effect is reduced (enhanced) in the presence of collective
flow in the positive (negative) jet direction, respectively. The
gluon formation factor must be averaged over the longitudinal target
profile, which is defined as $\left\langle \cdots\right\rangle=\int
dz \rho(z)\cdots$. We take the target distribution as an exponential
Gaussian form $\rho(z)=\exp(-z/L_e)/L_e$ with $L_e=L/2$, the gluon
formation factor can be deduced as
\begin{eqnarray}
 \label{LPMfactor}
 \left\langle Re(1{-}e^{\frac{i\omega_1z_{_{10}}}{1{-}v_z}})\right\rangle
 &=&
\frac{2}{L}\int_0^{\infty}d  z_{_{10}} e^{-{2z_{_{10}}}/{L}}
(1{-}e^{{\frac{i\omega_1z_{_{10}}}{1{-}v_z}}}) \nonumber \\
&=&\frac{(\mathbf{k_{\perp}}-\mathbf{q_{\perp}})^4L^2}{16x^2E^2(1-v_z)^2
+(\mathbf{k_{\perp}}-\mathbf{q_{\perp}})^4L^2}\, .
\end{eqnarray}
After integration over $\mathbf{k_{\perp}}$ and
$\mathbf{q_{\perp}}$, it can be obtained that gluon formation factor
is $1/(1-v_z)$ times as the static case. So that considering the flow along the jet direction,
 it increases gluon formation factor and decreases LPM effect. However,
square of radiation amplitude is decreased by the collective flow as
can be seen from Eq. (\ref{probafirstord}).

The jet energy loss can be divided into two parts. The
zero-temperature part corresponds to the radiation induced by
rescattering without detailed balance effect and can be expressed as
\begin{eqnarray}
 \label{energyloss1}
\Delta E^{(1)}_{rad}&=&\int d\omega \omega {{dP^{(1)}}\over
d\omega}\Big|_{T=0}
\nonumber \\
&=&\frac{\alpha_s C_R}{\pi}\frac{L}{l_g} E \int dx \int {{d{\bf
k}_{\perp}^2}\over {\bf k}_{\perp}^2} \int {d^2{\bf q}_{\perp}}
|{\bar v}({\bf q}_{\perp})|^2 P(x)(1-v_z)^2{{{\bf k}_{\perp}\cdot
{\bf q}_{\perp}} \over
      {\left({\bf k}_{\perp} - {\bf q}_{\perp}\right)^2}}
      \left\langle
      Re(1{-}e^{\frac{i\omega_1z_{10}}{1{-}v_z}})\right\rangle,\nonumber
      \\
\end{eqnarray}
where $l_g=C_R l/C_A$ is the mean-free path of the gluon.

The temperature-dependent part of energy loss induced by
rescattering to the first order in opacity comes from thermal
absorption with partial cancelation by stimulated emission, in the
presence of flow it can be written as
\begin{eqnarray}
 \label{eabs1}
   \Delta E^{(1)}_{abs}&=&\int d\omega \,\omega
     \left({{dP^{(1)}}\over d\omega} -{{dP^{(1)}}\over d\omega}
       \Big|_{T=0}\right) \nonumber \\
       &=&\frac{\alpha_s C_R}{\pi}\frac{L}{l_g} E \int dx \int {{d{\bf
k}_{\perp}^2}\over {\bf k}_{\perp}^2} \int {d^2{\bf q}_{\perp}}
|{\bar v}({\bf q}_{\perp})|^2 N_g(xE) (1-v_z)^2{{{\bf
k}_{\perp}\cdot {\bf q}_{\perp}} \over
      {\left({\bf k}_{\perp} - {\bf q}_{\perp}\right)^2}}
\nonumber \\
&&
      \big[ P({-}x)\left\langle
      Re(1{-}e^{\frac{i\omega_1z_{10}}{1{-}v_z}})\right\rangle
   -P(x)\left\langle
      Re(1-e^{\frac{i\omega_1z_{10}}{1-v_z}})\right\rangle\theta(1-x)\big],
 \end{eqnarray}
where $|{\bar v}({\bf q}_{\perp})|^2$ is defined as the normalized
distribution of momentum transfer from the scattering centers,
 \begin{eqnarray}
 \label{vbar}
  &&|{\bar v}({\bf q}_{\perp})|^2 \equiv {1\over \sigma_{el}}
  {d^2\sigma_{el}\over d^2{\bf q}_{\perp}}=
  {1\over \pi}{\mu^2_{eff}\over ({\bf q}_{\perp}^2+\mu^2)^2}\, ,
 \\
   &&{1\over \mu^2_{eff}} = {1\over \mu^2}-{1\over
   q_{\perp max}^2+\mu^2}\, ,\quad q_{\perp max}^2\approx 3E\mu\, .
 \end{eqnarray}

To obtain a simple analytic result, we take the kinematic boundaries
limit $q_{\perp max}\rightarrow\infty$, the angular integral can be
carried out by partial integration. In the limit of $EL \gg 1$ and
$E\gg \mu$,  we obtain the approximate asymptotic behavior of the
energy loss,
 \begin{equation}
 \label{elossem3}
   {\Delta E_{rad}^{(1)}\over E}{=}
   (1{-}v_z){{\alpha_s C_R \mu^2 L^2}\over 4\lambda_gE}
   \left[\ln{2E\over \mu^2L} {-}0.048\right]{+}\mathcal{O}(|\mathbf{v}|^2)\, ,
 \end{equation}
 \begin{eqnarray}
 \label{elossab3}
 {\Delta E_{abs}^{(1)}\over E}=-
   {{\pi\alpha_s C_R}\over 3} {{LT^2}\over {\lambda_g
   E^2}} \left[
   \ln{{\mu^2L}\over T} -1+\gamma_{\rm E}-{{6\zeta^\prime(2)}\over\pi^2}
\right]+\mathcal{O}(|\mathbf{v}|^2)\, .
 \end{eqnarray}
Although the collective flow reduces the square of radiation
amplitude when $v_z>0$, it increases the gluon distribution function
so that the energy gain to the first order in opacity is nearly the
same as that in the static medium. The collective flow hardly change
the energy gain to the first order in opacity.

Our analytic result implies, to the first order in opacity, the
emitted energy loss is changed by a factor $(1-v_z)$ with collective
flow. In addition, the collective flow increases the energy gain to
the zeroth order in opacity with a factor $(1+v_z)^2$ and hardly
change the energy gain to the first order in opacity.
Our results is
consistent with GLV static potential results when the velocity of
the collective flow goes to zero.

\begin{figure}
\begin{center}
\includegraphics[width=80mm]{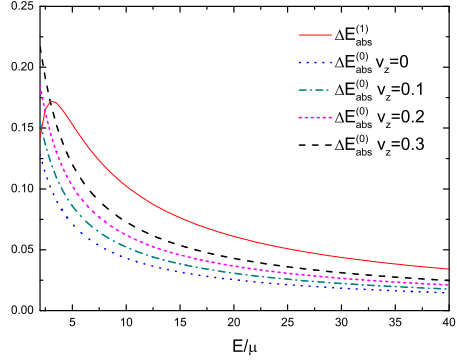}
\end{center}
\vspace{-12pt} \caption{\label{fig:eabs}
  The energy gain via gluon absorption with
rescattering for $v_z=0,0.1,0.2,0.3$ and without rescattering as
functions of $E/\mu$.  }
\end{figure}

Shown in Fig. \ref{fig:eabs} is the energy gain via gluon absorption with
rescattering to the first order in opacity for $v_z=0,0.1,0.2,0.3$ and without rescattering as
functions of $E/\mu$. For comparison, we take the same values for
the medium thickness, the mean free path, and the Debye screen mass
as in Refs.\cite{GLV} and \cite{Wang:2001sf}. The energy gain
without rescattering at very small $E/\mu$ region is larger than
that with rescattering if $v_z>0.2$. This is different with the
static case. But at smaller flow velocity or at higher jet energies,
the energy gain without rescattering becomes smaller than that with
rescattering.

\begin{figure}
\begin{center}
\includegraphics[width=80mm]{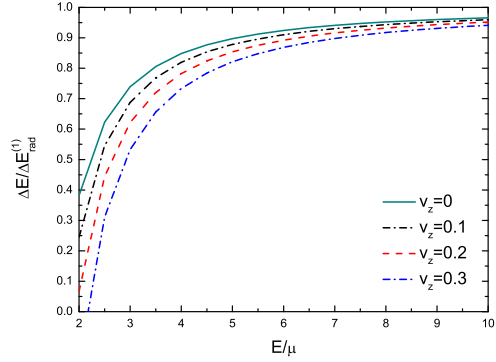}
\end{center}
\vspace{-12pt} \caption{\label{fig:ratioenergy}
  The ratio of effective parton energy loss with ($\Delta E=\Delta E_{rad}^{(1)}-\Delta E_{abs}^{(1)}-\Delta E_{abs}^{(0)}$) and without ($\Delta E=\Delta E_{rad}^{(1)}$)
absorption as a function of $E/\mu$ with collective flow velocity
$v_z=0,0.1,0.2,0.3$ in the positive jet direction.  }
\end{figure}

Shown in Fig.~\ref{fig:ratioenergy} is the ratio of the calculated radiative energy loss
with ($\Delta E=\Delta E_{rad}^{(1)}-\Delta E_{abs}^{(1)}-\Delta
E_{abs}^{(0)}$) and without ($\Delta E=\Delta E_{rad}^{(1)}$)
thermal absorption as functions of $E/\mu$
with collective flow velocity $v_z=0,0.1,0.2,0.3$ in the positive
jet direction. One sees that considering collective flow along jet
direction, this ratio decreases up to 40\% less than than that in
the static case. Hydrodynamics tells us that flow velocity in the
QGP medium is up to 0.3-0.4\cite{Hirano}. All these imply that for
the intermediate energy jet, such as mini-jet, the gluon absorption
shall be considered instead of ignoring the gluon absorption effect.

\subsection{Second Order in Opacity Correction}

It was shown by GLV \cite{Wiedua} that in the static medium, the induced
gluon radiation intensity is dominated by the first order opacity contribution, higher
order correction contribute little to the radiative energy loss. Here we will analyze the
seconder order in opacity correction with considering collective flow in the moving medium.

At the second order in opacity, consider the jet has the simplest
case of two consecutive elastic rescatterings. The radiation
amplitude
 \begin{eqnarray}
 \label{rad2}
   M^{(2)} &\propto& (-i)(2p-q_1-2q_2)_{\mu}A^{\mu}(-i)(2p-q_2)_{\nu}A^{\nu} \nonumber \\
           &\propto& 2\pi\delta(q_1^0{-}\mathbf{v}\cdot\mathbf{q}_1)
e^{-i\mathbf{q}_1\cdot \mathbf{x}_1}
T_{a_1}(j)T_{a_1}(i)(2E+\mathbf{v}\cdot \mathbf{q}_1) \nonumber
\\
&& \times 2\pi\delta(q_2^0{-}\mathbf{v}\cdot\mathbf{q}_2)
e^{-i\mathbf{q}_2\cdot \mathbf{x}_2}
T_{a_2}(j)T_{a_2}(i)(2E+\mathbf{v}\cdot \mathbf{q}_2)R^{(2)}
           \, ,
 \end{eqnarray}
 where $R^{(2)}=(1-v_z)^2\tilde{v}(\mathbf{q}_1)\tilde{v}(\mathbf{q}_2)$, which is changed by
 the collective flow with a factor $(1-v_z)^2$.

\begin{figure}
\begin{center}
\includegraphics[width=80mm]{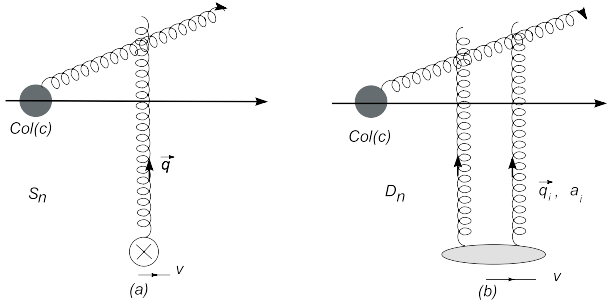}
\end{center}
\vspace{-12pt} \caption{\label{fig:Feydia} Diagrammatic representation of
 a "direct" interaction of $S_n$ and a double Born "virtual" interaction $D_n$}
\end{figure}

The inclusive gluon distribution to the second order in opacity is a sum of $7^2$ direct plus
$2\times86$ virtual cut diagrams. It is useful to write the sum of amplitudes in a certain class
of diagrams\cite{Wiedua}. There are two basic iteration steps to construct the inclusive distribution
of gluon. The first one represent the addition of a "direct" interaction of $S_n$ that changes the
color or momentum of the target parton with flow velocity $\vec{v}$ located at $\tilde{x_i}$ as illustrated in Fig.~(\ref{fig:Feydia}a).
The second one corresponds to a double Born "virtual" interaction $D_n$ that leaves both the color and
momentum of the target parton unchanged as in Fig.~(\ref{fig:Feydia}b). As discussed in the last section, $R^{S_1}$
and $R^{D_1}$ correspond to the first order in opacity. To the second order in opacity, four
new classes of factorized radiation amplitude emerge. They are readily derived from $R^{S_n}$ and $R^{D_n}$ with $n\leq 2$.

The first class of the factorized radiation amplitude include two single rescattering, which can be written as,
\begin{eqnarray}
 \label{rss}
   R^{(S_1 S_2)}&=&2ig\cdot {\mathbf \epsilon}_{\perp}(1-v_z)^2 \Bigr[
   \Bigr({\bf H} e^{\frac{i\omega_0z_{0}}
   {1-v_z}}T_{a_2}T_{a_1}T_{c}
    +{\bf C}_1
    e^{\frac{i(\omega_0z_{1}-\omega_1z_{10})}{1-v_z}}T_{a_2}[T_{c},T_{a_1}]
 \nonumber \\
    &&
     +{\bf C}_2
    e^{\frac{i(\omega_0z_{2}-\omega_2z_{20})}{1-v_z}}T_{a_1}[T_{c},T_{a_2}]+{\bf
    B}_1
    e^{\frac{i\omega_0z_{1}}{1-v_z}}T_{a_2}[T_{c},T_{a_1}]
    \nonumber \\
    &&
    +{\bf B}_2 e^{\frac{i\omega_0z_{2}}{1-v_z}}[T_{c},T_{a_2}]T_{a_1}+{\bf
    C}_{(12)}
    e^{\frac{i(\omega_0z_{2}-\omega_2z_{21}-\omega_{(12)}z_{10})}{1-v_z}}[[T_{c},T_{a_2}],T_{a_1}]
    \nonumber \\
    &&
    +{\bf
    B}_{2(12)}
    e^{\frac{i(\omega_0z_{2}-\omega_2z_{21})}{1-v_z}}[[T_{c},T_{a_2}],T_{a_1}]\Bigr)
    \nonumber \\
    &&
    +2v_z\cdot\Bigr(-{\bf H} e^{\frac{i\omega_0z_{0}}
   {1-v_z}}T_{a_2}T_{a_1}T_{c}-{\bf H} e^{\frac{i\omega_0z_{1}}
   {1-v_z}}T_{a_2}[T_{c},T_{a_1}]+{\bf H} e^{\frac{i\omega_0z_{2}}
   {1-v_z}}T_{a_2}T_{c}T_{a_1}
   \nonumber \\
   &&
   -{\bf
    C}_{(12)}
    e^{\frac{i(\omega_0z_{2}-\omega_2z_{21})}{1-v_z}}[[T_{c},T_{a_2}],T_{a_1}]
    +{\bf
    C}_{(12)}
    e^{\frac{i(\omega_0z_{2}-\omega_2z_{21}-\omega_{(12)}z_{10})}{1-v_z}}[[T_{c},T_{a_2}],T_{a_1}]\Bigr)
    \Bigr]
     ,
\end{eqnarray}
where
\begin{eqnarray}
 \label{shorthand}
   \omega_{(ij\dots)}=\frac{({\bf k}_{\perp}{-}{\bf q}_{i\perp}{-}{\bf q}_{j\perp}{-}\dots)^2}
      {2\omega}\, &,&
 \quad
   {\bf C}_{(ij\dots)}=\frac{{\bf k}_{\perp}{-}{\bf q}_{i\perp}{-}{\bf q}_{j\perp}{-}\dots}
      {({{\bf k}_{\perp}{-}{\bf q}_{i\perp}}{-}{\bf q}_{j\perp}{-}\dots)^2}\, ,
   \\
   {\bf B}_{(i_1 i_2\cdots i_m)(j_1 j_2\cdots j_n)}&=&{\bf C}_{(i_1 i_2\cdots
   i_m)}{-}{\bf C}_{(j_1 j_2\cdots
   j_n)}\, ,
 \end{eqnarray}
and $z_{ij}=z_i - z_j$.

Secondly, the radiation amplitude of a single hit followed by a double
Born scattering can be expressed as
\begin{eqnarray}
 \label{rsd}
   R^{(S_1 D_2)}&=&2ig\cdot {\mathbf \epsilon}_{\perp}(1-v_z)^3 \Bigr[
   \Bigr(-\frac{C_R+C_A}{2}{\bf H} e^{\frac{i\omega_0z_{0}}
   {1-v_z}}T_{a_1}T_{c}
    -\frac{C_R+C_A}{2}{\bf C}_1
    e^{\frac{i(\omega_0z_{1}-\omega_1z_{10})}{1-v_z}}[T_{c},T_{a_1}]
 \nonumber \\
    &&
     -{\bf C}_2
    e^{\frac{i(\omega_0z_{2}-\omega_2z_{20})}{1-v_z}}T_{a_2}T_{a_1}[T_{c},T_{a_2}]
    -\frac{C_R+C_A}{2}{\bf
    B}_1
    e^{\frac{i\omega_0z_{1}}{1-v_z}}[T_{c},T_{a_1}]+\frac{C_A}{2}{\bf
    B}_2
    e^{\frac{i\omega_0z_{2}}{1-v_z}}T_{c}T_{a_1}
    \nonumber \\
    &&
    -{\bf
    C}_{(12)}
    e^{\frac{i(\omega_0z_{2}-\omega_2z_{21}-\omega_{(12)}z_{10})}{1-v_z}}T_{a_2}[[T_{c},T_{a_2}],T_{a_1}]
    \nonumber \\
    &&
    -{\bf
    B}_{2(12)}
    e^{\frac{i(\omega_0z_{2}-\omega_2z_{21})}{1-v_z}}T_{a_2}[[T_{c},T_{a_2}],T_{a_1}]\Bigr)
    \nonumber \\
    &&
    +2v_z\cdot\Bigr( -\frac{C_A-C_R}{2}{\bf H} e^{\frac{i\omega_0z_{0}}
   {1-v_z}}T_{a_1}T_{c}
    +\frac{C_R}{2}{\bf H}
    e^{\frac{i\omega_0z_{1}}{1-v_z}}[T_{c},T_{a_1}]
 \nonumber \\
    &&
     -\frac{C_R-C_A}{2}{\bf H}
    e^{\frac{i\omega_0z_{2}}{1-v_z}}T_{c}T_{a_1}
    -\frac{C_A}{2}{\bf
    H}
    e^{\frac{i\omega_0z_{1}}{1-v_z}}[T_{c},T_{a_1}]+{C_A}{\bf
    C}_1
    e^{\frac{i\omega_0z_{1}}{1-v_z}}[T_{c},T_{a_1}]
    \nonumber \\
    &&
    -{C_A}{\bf
    C}_{1}
    e^{\frac{i(\omega_0z_{1}-\omega_1z_{10})}{1-v_z}}[T_{c},T_{a_1}]
    \nonumber \\
    &&
    +{\bf
    C}_{(12)}
    e^{\frac{i(\omega_0z_{2}-\omega_2z_{21})}{1-v_z}}T_{a_2}[[T_{c},T_{a_2}],T_{a_1}]
    \nonumber \\
    &&
    -{\bf
    C}_{(12)}
    e^{\frac{i(\omega_0z_{2}-\omega_2z_{21}-\omega_{(12)}z_{10})}{1-v_z}}T_{a_2}[[T_{c},T_{a_2}],T_{a_1}]
      \Bigr)
    \Bigr]
     .
\end{eqnarray}

Interchanging the sequence of the single hit and the double Born
interaction in the above process, one would find a new radiation
amplitude:
\begin{eqnarray}
 \label{rds}
   R^{(D_1 S_2)}&=&2ig\cdot {\mathbf \epsilon}_{\perp}(1-v_z)^3 \Bigr[
   \Bigr(-\frac{C_R+C_A}{2}{\bf H} e^{\frac{i\omega_0z_{0}}
   {1-v_z}}T_{a_2}T_{c}
    +\frac{C_A}{2}{\bf C}_1
    e^{\frac{i(\omega_0z_{1}-\omega_1z_{10})}{1-v_z}}T_{a_2}T_{c}
 \nonumber \\
    &&
     -\frac{C_R+C_A}{2}{\bf C}_2
    e^{\frac{i(\omega_0z_{2}-\omega_2z_{20})}{1-v_z}}[T_{c},T_{a_2}]
    +\frac{C_A}{2}{\bf
    B}_1
    e^{\frac{i\omega_0z_{1}}{1-v_z}}T_{a_2}T_{c}-\frac{C_R}{2}{\bf
    B}_2
    e^{\frac{i\omega_0z_{2}}{1-v_z}}[T_{c},T_{a_2}]
    \nonumber \\
    &&
    +\frac{C_A}{2}{\bf
    C}_{(12)}
    e^{\frac{i(\omega_0z_{2}-\omega_2z_{21}-\omega_{(12)}z_{10})}{1-v_z}}[T_{c},T_{a_2}]
    +\frac{C_A}{2}{\bf
    B}_{2(12)}
    e^{\frac{i(\omega_0z_{2}-\omega_2z_{21})}{1-v_z}}[T_{c},T_{a_2}]\Bigr)
    \nonumber \\
    &&
    +2v_z\cdot\Bigr( -\frac{C_A-C_R}{2}{\bf H} e^{\frac{i\omega_0z_{0}}
   {1-v_z}}T_{a_2}T_{c}
    +\frac{C_A}{2}{\bf H}
    e^{\frac{i\omega_0z_{1}}{1-v_z}}T_{a_2}T_{c}
 \nonumber \\
    &&
     +(C_A{\bf C}_{2}-\frac{C_A}{2}{\bf C}_{(12)})
    e^{\frac{i(\omega_0z_{2}-\omega_2z_{21})}{1-v_z}}[T_{c},T_{a_2}]
    -\frac{C_R}{2}{\bf
    H}
    e^{\frac{i\omega_0z_{2}}{1-v_z}}T_{a_2}T_{c}
    \nonumber \\
    &&
    -{C_A}{\bf
    C}_2
    e^{\frac{i(\omega_0z_{2}-\omega_2z_{20})}{1-v_z}}[T_{c},T_{a_2}]
    +\frac{C_A}{2}{\bf
    C}_{(12)}
    e^{\frac{i(\omega_0z_{2}-\omega_2z_{21}-\omega_{(12)}z_{10})}{1-v_z}}[T_{c},T_{a_2}]
      \Bigr)
    \Bigr]
     ,
\end{eqnarray}

Then, if a jet parton encounters two double Born interaction, its
radiation amplitude will be
\begin{eqnarray}
 \label{rdd}
   R^{(D_1 D_2)}&=&2ig\cdot {\mathbf \epsilon}_{\perp}(1-v_z)^4 \Bigr[
   \Bigr(\frac{(C_R+C_A)^2}{4}{\bf H} e^{\frac{i\omega_0z_{0}}
   {1-v_z}}T_{c}
    -\frac{C_A(C_R+C_A)}{4}{\bf C}_1
    e^{\frac{i(\omega_0z_{1}-\omega_1z_{10})}{1-v_z}}T_{c}
 \nonumber \\
    &&
     -\frac{C_A(C_R+C_A)}{4}{\bf C}_2
    e^{\frac{i(\omega_0z_{2}-\omega_2z_{20})}{1-v_z}}T_{c}
    -\frac{C_A(C_R+C_A)}{4}{\bf
    B}_1
    e^{\frac{i\omega_0z_{1}}{1-v_z}}T_{c}-\frac{C_A C_R}{4}{\bf
    B}_2
    e^{\frac{i\omega_0z_{2}}{1-v_z}}T_{c}
    \nonumber \\
    &&
    +\frac{C_A^2}{4}{\bf
    C}_{(12)}
    e^{\frac{i(\omega_0z_{2}-\omega_2z_{21}-\omega_{(12)}z_{10})}{1-v_z}}T_{c}
    +\frac{C_A^2}{4}{\bf
    B}_{2(12)}
    e^{\frac{i(\omega_0z_{2}-\omega_2z_{21})}{1-v_z}}T_{c}\Bigr)
    \nonumber \\
    &&
    +2v_z\cdot\Bigr( -\frac{(C_A-C_R)^2}{4}{\bf H} e^{\frac{i\omega_0z_{0}}
   {1-v_z}}T_{c}
    +{C_A^2}{\bf H} e^{\frac{i\omega_0z_{0}}
   {1-v_z}}T_{c}
   \nonumber \\
   &&
   -\frac{C_A(C_R+C_A)}{4}{\bf H}
    e^{\frac{i\omega_0z_{1}}{1-v_z}}T_{c}
       -\frac{C_A^2}{2}{\bf B}_{1}
    e^{\frac{i\omega_0z_{1}}{1-v_z}}T_{c}
 \nonumber \\
    &&
     +\frac{C_A^2}{4}{\bf C}_{(12)}
    e^{\frac{i(\omega_0z_{2}-\omega_2z_{21}-\omega_{(12)}z_{10})}{1-v_z}}T_{c}
    -\frac{C_A^2}{2}{\bf
    C}_2
    e^{\frac{i(\omega_0z_{2}-\omega_2z_{20})}{1-v_z}}T_{c}
    \nonumber \\
    &&
    -\frac{C_A^2}{2}{\bf
    C}_1
    e^{\frac{i(\omega_0z_{1}-\omega_1z_{10})}{1-v_z}}T_{c}
    +\frac{C_A^2}{4}{\bf B}_{2(12)}
    e^{\frac{i(\omega_0z_{2}-\omega_2z_{21})}{1-v_z}}T_{c}
    \nonumber \\
    &&
    +\frac{C_A^2}{4}{\bf
    C}_2
    e^{\frac{i(\omega_0z_{2}-\omega_2z_{21})}{1-v_z}}T_{c}
      -\frac{C_R(C_A-C_R)}{4}{\bf H}
    e^{\frac{i\omega_0z_{2}}{1-v_z}}T_{c}\Bigr)
    \Bigr]
     ,
\end{eqnarray}
We can see clearly that these amplitudes are influenced by
collective flow.

We recall the three amplitudes $R^{(0)}$, $R^{(S_1)}$ and $R^{(D_1)}$ at the zeroth and
first order in opacity respectively,  to obtain the radiation probability to the second order
in opacity,
\begin{eqnarray}
\label{radpropbility2}
  \frac{{dP}^{( 2 )}}{d  \omega}&=&\frac{C_{R}
  \alpha_{s}}{2 \pi^{2}} \left( \frac{L}{l_{g}} \right)^{2} \int \frac{d x}{x} \int
  d^{2} \vec{k}_{\perp} \int d^{2}
 \vec{q}_{1 \perp} \int d^{2} \vec{q}_{2 \perp} P \left( \frac{\omega}{E} \right)
 J_{eff}^{(2)}(k_{\perp},\mathbf{q}_{\perp})\nonumber\\
     &&
     \times\Big[\left(1{+}N_g(xE)\right)\delta(\omega{-}xE)\theta(1{-}x)
      {+}N_g(xE)\delta(\omega{+}xE)\Big].
\end{eqnarray}
Here the
``emission current'' to the second order in opacity $J_{eff}^{(2)}(k_{\perp},\mathbf{q}_{\perp})$ is,
\begin{eqnarray}
 \label{emissioncurrent2}
J_{eff}^{(2)}(k_{\perp},\mathbf{q}_{\perp})
&=&|R^{(2)}|^2\Bigr\langle Tr\Bigr[ |R^{(S_1 S_2)}|^2+2Re(R^{(S_1
D_2)}R^{(S_1)\dagger}) \nonumber \\
&&+2Re(R^{(D_1 S_2)}R^{(S_2)\dagger})+2Re(R^{(D_1)}R^{(D_1)\dagger})
+2Re(R^{(D_1 D_2)}R^{(0)\dagger})\Bigr]\Bigr\rangle
\\
& \approx & (1-v_z)^8 |{\bar v}({\bf q}_{1\perp})|^2 |{\bar v}({\bf
q}_{2\perp})|^2
\Bigr[\Bigr(2\textbf{C}_{1}\cdot\textbf{B}_{1}\langle
Re(1-e^{\frac{i\omega_1z_{10}}{1-v_z}})\rangle\nonumber \\
&& -2\textbf{C}_{12}\cdot\textbf{B}_{2(12)}\langle
Re(1-e^{\frac{i\omega_{12}z_{10}}{1-v_z}})\rangle \nonumber \\
&& +2\textbf{C}_{2}\cdot\textbf{B}_{2}(\langle Re(e^{\frac{i\omega_2
z_{21}}{1-v_z}})\rangle-\langle Re(e^{\frac{i\omega_2(
z_{10}+z_{21})}{1-v_z}})\rangle)\nonumber \\
&&  -2\textbf{C}_{12}\cdot\textbf{B}_{2}(\langle
Re(e^{\frac{i\omega_2 z_{21}}{1-v_z}})\rangle-\langle
Re(e^{\frac{i(\omega_{12}
z_{10}+\omega_2 z_{21})}{1-v_z}})\rangle)\Bigr)\nonumber
\\
&& +v_{z}\Bigr(4\textbf{C}_{1}\cdot\textbf{B}_{1}\langle
Re(1-e^{\frac{i\omega_1z_{10}}{1-v_z}})\rangle-4\textbf{C}_{12}\cdot\textbf{B}_{2(12)}\langle
Re(1-e^{\frac{i\omega_{12}z_{10}}{1-v_z}})\rangle\nonumber \\
&& +(2\textbf{C}_{2}\cdot\textbf{B}_{2}-2\textbf{C}_{2}^{2})(\langle
Re(e^{\frac{i\omega_2 z_{21}}{1-v_z}})\rangle-\langle
Re(e^{\frac{i\omega_2(z_{10}+z_{21})}{1-v_z}})\rangle)\nonumber \\
&&  -(2\textbf{C}_{12}\cdot\textbf{B}_{2}-2\textbf{C}_{2}\cdot
\textbf{C}_{12})(\langle Re(e^{\frac{i\omega_2
z_{21}}{1-v_z}})\rangle-\langle Re(e^{\frac{i(\omega_{12}
z_{10}+\omega_2 z_{21})}{1-v_z}})\rangle) \Bigr)\Bigr].
\end{eqnarray}
The gluon formation factors lead to non-Abelian LPM effect. To the
second order in opacity, the gluon formation factors are also averaged
over the target profile, which is taken as $\rho(z)=\exp(-z/L_e)/L_e$ with $L_e=L/3$.
This converts the gluon formation factors into simple Lorentzian factors,
\begin{eqnarray}\label{LPM2}
<Re[\mathbf{Exp}({i\frac{\sum\limits_{j=k}^{m} \omega_{(k,\ldots,j,\ldots)}\triangle z_k}{1-v_z}})]>&=&\int d \rho
Re[\mathbf{Exp}({i\frac{\sum\limits_{j=k}^{m} \omega_{(k,\ldots,j,\ldots)}\triangle z_k}{1-v_z}})] \nonumber
\\
&=& Re \prod \limits_{j=k}^{m}\frac{1}{1+i\frac{\omega_{(k,\ldots,j,\ldots)}}{1-v_z}\frac{L}{3}}
\end{eqnarray}
where $m$ is the subscript of $\textbf{B}$ in the same term, and $k\leq j\leq m$, $0\leq k,j,m \leq 2$.
As can be seen here, to the second order in opacity, the collective flow also reduces the LPM effect if considering
flow along the jet direction.

The radiative energy loss to the second order in opacity corresponds to the zero-temperature part, it can be
expressed as,
\begin{eqnarray}
 \label{energyloss2}
\Delta E^{(2)}_{rad}&=&\int d\omega \omega {{dP^{(2)}}\over
d\omega}\Big|_{T=0}
\nonumber \\
&=&\frac{C_{R}
  \alpha_{s}}{2 \pi^{2}} \left( \frac{L}{l_{g}} \right)^{2} E \int dx \int
  d^{2} \vec{k}_{\perp} \int d^{2}
 \vec{q}_{1 \perp} \int d^{2} \vec{q}_{2 \perp} P \left( \frac{\omega}{E} \right)
 J_{eff}^{(2)}(k_{\perp},\mathbf{q}_{\perp}),
\end{eqnarray}

\begin{figure}
\begin{center}
\includegraphics[width=85mm]{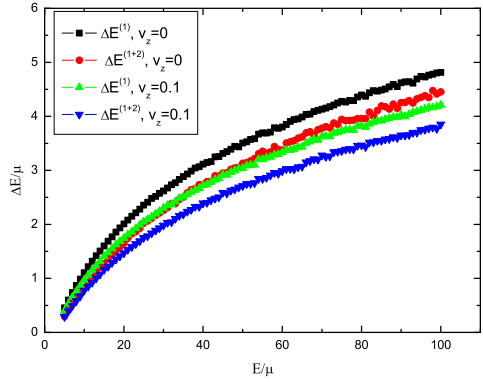}
\end{center}
\vspace{-12pt} \caption{\label{fig:radienergy12}
  The radiative jet energy loss with and without collective
flow as a functions of $E/\mu$.  }
\end{figure}

The numerical result of radiative energy loss as a function of $E/\mu$ is shown in
Fig.(~\ref{fig:radienergy12}). We take the same values for the medium thickness, the mean free path,
and the Debye screen mass as in Refs.\cite{GLV} and \cite{Wang:2001sf} for comparison. It shows that
in the presence of collective flow, the second order in opacity correction is relatively small as compared to
the first  order in opacity. The contribution of the first order is dominant.

In the moving medium, to the second order in opacity, the energy gain via
gluon absorption corresponding to the temperature-dependent part of energy
loss also comes from the partial cancelation by stimulated emission, it
can be written as,
\begin{eqnarray}
 \label{eabs1}
   \Delta E^{(2)}_{abs}&=&\int d\omega \,\omega
     \left({{dP^{(2)}}\over d\omega} -{{dP^{(2)}}\over d\omega}
       \Big|_{T=0}\right) \nonumber \\
       &=&\frac{C_{R}
  \alpha_{s}}{2 \pi^{2}} \left( \frac{L}{l_{g}} \right)^{2} E \int dx \int
  d^{2} \vec{k}_{\perp} \int d^{2}
 \vec{q}_{1 \perp} \int d^{2} \vec{q}_{2 \perp}
 J_{eff}^{(2)}(k_{\perp},\mathbf{q}_{\perp})\nonumber \\
 &&\times N_g(xE)\big[ P({-}x)-P(x)\theta(1-x)\big].
 \end{eqnarray}

We calculated the numerical result and find that the energy gain to the second order in opacity is little
as compared to the energy gain to the first order in opacity. So that it is negligible to the total effective energy loss.


\section{Conclusion}

In summary, we have derived a new potential for the interaction of a
hard jet with the parton target. It can be used to study the jet
quenching phenomena in the presence of collective flow of the
quark-gluon medium. With this new potential, we have investigated
the effect of collective flow on jet energy loss with detailed
balance. Collective flow along the jet direction leads to decreased
LPM effect and the square of radiation amplitude. To the zeroth order
in opacity, the energy gain  without rescattering is $(1+v_z)^2$ times as
in the static medium. To the first order in opacity, the gluon emission
energy loss is $(1-v_z)$ times as that in the static
medium, but the energy gain is nearly the same as that in the static medium.
To the second order in opacity, both the radiative energy loss and the energy gain
are relatively small as compared to the first order.  All these imply the
collective flow along the jet direction decreases gluon emission
energy loss, increases energy gain, so that the total effective
energy loss is decreased by collective flow. All these lead to that
for the intermediate energy jet, such as mini-jet, the gluon
absorption shall be considered. Compared to calculations for a
static medium, our results will affect the suppression of high $p_T$
hadron spectrum and anisotropy parameter $v_2$ in high-energy
heavy-ion collisions. Our new potential can also be used for heavy
quark energy loss calculation and will alter the dead cone effect of
heavy quark jets. Our results shall have implications for
comparisons between theory and experiment in the future.

\section{Acknowledgments}
We thank Fuqiang Wang, Xin-Nian Wang and Ulrich Heinz for helpful
discussions and comments. This work was supported by NSFC of China
under Grants No. 11205024, No. 11221504 and No. 10825523, the Major
 State Basic Research Development Program in China (No. 2014CB845404),
  and Ministry of Education of China
under the Doctoral Grants No. 20120041120043.


\begin{thebibliography}{99}


\bibitem{Lorstad}
T. Cs\"{o}g\"{o}, B. Lorstad, Phys. Rev. C {\bf 54}, (1996) 1390.

\bibitem{Voloshin1}
 S.~A.~Voloshin, W.~E.~Cleland, Phys. Rev. C {\bf 53}, (1996) 896.

\bibitem{Wiedemann1}
 U.A. Wiedemann, Phys. Rev. C {\bf 57}, (1997) 266.

\bibitem{Voloshin2}
S. Voloshin, R. Lednicky, S. Panitkin, Nu Xu, Phys. Rev. Lett. {\bf
79}, (1997) 4766.

\bibitem{phenix}
  K.~Adcox {\it et al.},
  Phys.\ Rev.\ Lett. {\bf 88}, (2001) 022301;
  C.~Adler {\it et al.},
  Phys.\ Rev.\ Lett. {\bf 89}, (2002) 202301.


\bibitem{star} C.~Adler {\it et al.},
  Phys.\ Rev.\ Lett. {\bf 90}, (2003) 082302.


\bibitem{GW94}
M.~Gyulassy, X.-N.~Wang,
Nucl.\ Phys.\ B {\bf 420}, (1994) 583;
X.-N.~Wang, M.~Gyulassy, M.~Plumer,
Phys.\ Rev.\ D {\bf 51}, (1995) 3436.


\bibitem{BDPMS}
R.~Baier, Y.~L.~Dokshitzer, S.~Peigne, D.~Schiff,
Phys.\ Lett.\ B {\bf 345}, (1995) 277;
R.~Baier, Y.~L.~Dokshitzer, A.~H.~Mueller, S.~Peigne, D.~Schiff,
Nucl.\ Phys.\ B {\bf 484}, (1997) 265.

\bibitem{Zakharov}
B.~G.~Zakharov,
JETP Lett. {\bf 63}, (1996) 952;
JETP Lett. {\bf 65}, (1997) 615.

\bibitem{GLV}
M.~Gyulassy, P.~Levai, I.~Vitev,
Phys.\ Rev.\ Lett. {\bf 85}, (2000) 5535;
Nucl.\ Phys.\ B {\bf 594}, (2001) 371.

\bibitem{Wiedua}
U.~A.~Wiedemann,
Nucl.\ Phys.\ B {\bf 588}, (2000) 303.


\bibitem{GuoW}
X.~Guo, X.-N.~Wang,
Phys.\ Rev.\ Lett. {\bf 85}, (2000) 3591.


\bibitem{BDMS}
R.~Baier, Y.~L.~Dokshitzer, A.~H.~Mueller, D.~Schiff, Phys.\ Rev.\ C
{\bf 58}, (1998) 1706.

\bibitem{SW}
C.~A.~Salgado, U.~A.~Wiedemann, Phys. Rev. Lett. {\bf 89} (2002)
092303.

\bibitem{WW}
E.~Wang, X.-N.~Wang, Phys. Rev. Lett. {\bf 89} (2002) 162301.


\bibitem{ASWiedemann}
N. Armesto, C. A. Salgado, U.A. Wiedemann, Phys. Rev. Lett. {\bf 93}
(2004) 242301.

\bibitem{BMS}
R.~Baier, A.~H.~Mueller, D.~Schiff, Phys. Lett. B. {\bf 649} (2007)
147.

\bibitem{Liu}
H.~Liu, K.~Rajagopal, U.~A.~Wiedemann, JHEP {\bf 0703} (2007) 066.

\bibitem{Wang:2001sf}
E.~Wang, X.-N.~Wang, Phys.\ Rev.\  Lett. {\bf 87} (2001) 142301.

\bibitem{Hirano}
T. Hirano, Phys. Rev. C65, 011901(2001); T. Hirano and K. Tsuda,
Phys. Rev. C66, 054905 (2002); T. Hirano, U. Heinz, D. Kharzeev, R.
Lacey and Y. Nara, Phys. Lett. B636, 299 (2006); Phys. Rev. C77,
044909 (2008).


\end{thebibliography}
\end{document}